\documentclass[iop]{emulateapj}

\usepackage{color}
\usepackage{savesym}
\restoresymbol{SIXtablenum}{tablenum}

\usepackage{enumerate}

\definecolor{red}{rgb}{1.0,0.0,0.0}

\shorttitle{PDS 66 Polarimetry with GPI}
\shortauthors{Wolff et al. 2015}

\begin{document}

\title{The PDS 66 Circumstellar Disk as seen in Polarized Light with the Gemini Planet Imager}


\author{Schuyler  G. Wolff\altaffilmark{1}, Marshall Perrin\altaffilmark{2}, Maxwell A. Millar-Blanchaer\altaffilmark{3, 4}, Eric L. Nielsen\altaffilmark{5,6}, Jason Wang\altaffilmark{7}, Andrew Cardwell\altaffilmark{8, 9}, Jeffrey Chilcote\altaffilmark{4},  Ruobing Dong\altaffilmark{7}, Zachary H. Draper\altaffilmark{10,11}, Gaspard Duch{\^e}ne\altaffilmark{7,12}, Michael P. Fitzgerald\altaffilmark{13},  Stephen J. Goodsell\altaffilmark{14}, Carol A. Grady\altaffilmark{15}, James R. Graham\altaffilmark{7}, Alexandra Z. Greenbaum\altaffilmark{1}, Markus Hartung\altaffilmark{9}, Pascale Hibon\altaffilmark{9}, Dean C. Hines\altaffilmark{2},  Li-Wei Hung\altaffilmark{13}, Paul Kalas\altaffilmark{7}, Bruce Macintosh\altaffilmark{5}, Franck Marchis\altaffilmark{6},  Christian Marois\altaffilmark{11}, Laurent Pueyo\altaffilmark{2},  Fredrik T. Rantakyr{\"o}\altaffilmark{9}, Glenn Schneider\altaffilmark{16}, Anand Sivaramakrishnan\altaffilmark{2, 17}, and Sloane J.  Wiktorowicz\altaffilmark{18, 19}}

\affil{$^{1}$ Department of Physics and Astronomy, Johns Hopkins University, Baltimore, MD 21218, USA}
\email{swolff9@jh.edu}
\affil{$^{2}$ Space Telescope Science Institute, Baltimore, MD 21218, USA}
\affil{$^{3}$ Department of Astronomy \& Astrophysics, University of Toronto, Toronto, ON, M5S 3H4, Canada}
\affil{$^{4}$ Dunlap Institute for Astronomy and Astrophysics, University of Toronto, Toronto, ON,  M5S 3H4, Canada}
\affil{$^{5}$ Kavli Institute for Particle Astrophysics and Cosmology, Stanford University, Stanford, CA 94305, USA}
\affil{$^{6}$ SETI Institute, Carl Sagan Center, 189 Bernardo Avenue,  Mountain View, CA 94043, USA}
\affil{$^{7}$ Astronomy Department, University of California, Berkeley; Berkeley, CA 94720, USA}

\affil{$^{8}$ LBT Observatory, University of Arizona, 933 N. Cherry Ave, Room 552 Tucson, AZ 85721, U.S.A.}
\affil{$^{9}$ Gemini Observatory, Casilla 603, La Serena, Chile}
\affil{$^{10}$ University of Victoria, 3800 Finnerty Rd, Victoria, BC, V8P 5C2, Canada}
\affil{$^{11}$ National Research Council of Canada Herzberg, 5071 West Saanich Rd, Victoria, BC, V9E 2E7, Canada}
\affil{$^{12}$ Univ. Grenoble Alpes/CNRS, IPAG, F-38000 Grenoble, France}
\affil{$^{13}$ Department of Physics \& Astronomy, 430 Portola Plaza, University of California, Los Angeles, CA 90095, USA}
\affil{$^{14}$ Gemini Observatory, 670 N. A'ohoku Place, Hilo, HI 96720, USA}
\affil{$^{15}$Eureka Scientific, 2452 Delmer, Suite 100, Oakland, CA 96002, USA}
\affil{$^{16}$ Steward Observatory and the Department of Astronomy, University of Arizona, 933 North Cherry Avenue, Tucson, AZ 85721, USA}
\affil{$^{17}$ Department of Astrophysics, American Museum of Natural History, New York, NY 10024, USA}
\affil{$^{18}$ Department of Astronomy \& Astrophysics, University of California, Santa Cruz, CA 95064, USA}
\affil{$^{19}$ Remote Sensing Department, The Aerospace Corporation, El Segundo, CA 90245, USA}


\begin{abstract}
We present \textit{H} and \textit{K} band imaging polarimetry for the PDS 66 circumstellar disk obtained during the commissioning of the Gemini Planet Imager (GPI). Polarization images reveal a clear detection of the disk in to the $0.12 \arcsec$ inner working angle (IWA) in \textit{H} band, almost 3 times as close to the star as the previous \textit{HST} observations with NICMOS and STIS ($0.35 \arcsec$ effective IWA). The centro-symmetric polarization vectors confirm that the bright inner disk detection is due to circumstellar scattered light. A more diffuse disk extends to a bright outer ring centered at 80 AU. We discuss several physical mechanisms capable of producing the observed ring + gap structure. GPI data confirm enhanced scattering on the East side of the disk which is inferred to be nearer to us. We also detect a lateral asymmetry in the South possibly due to shadowing from material within the inner working angle. This likely corresponds to a temporally variable azimuthal asymmetry observed in \textit{HST}/STIS coronagraphic imaging. 

\end{abstract}


\section{Introduction}

Classical T Tauri stars (cTTS) with optically-thick, gas-rich protoplanetary disks provide valuable knowledge of the precedent conditions for planet formation. By comparing the observed intensity in scattered light to radiative transfer models,  we can infer the grain properties (size, density, composition) and the geometry at the surface of the disk \citep[e.g.][]{2007ApJ...654..595G, 2010A&A...518A..63M}.

\begin{deluxetable*}{cccccccl}[t!]
\tablecolumns{8}
\tabletypesize{\footnotesize}
\tablecaption{Gemini Planet Imager observations of PDS 66}
\tablehead{
 \colhead{} &
 \colhead{} & 
 \colhead{} & 
 \colhead{Exp.} & 
 \colhead{\# of} & 
 \colhead{Coronagraph} &
 \colhead{Field} & 
 \colhead{} \\ 
 \colhead{Date} & 
 \colhead{Mode} &
 \colhead{Band} &
 \colhead{Time (s)} &
 \colhead{Exposures} &  
 \colhead{Spot Size ('')} &
 \colhead{Rotation ($\degr$)} & 
 \colhead{Notes} 
  }
\startdata
2014 May 14 & Spectral     & $H$  & 59.6 & 10 & 0.246  &   3.5  &  0.5 -- 0.8 $\arcsec$ seeing, high winds\\
2014 May 15 & Polarization & $H$  & 59.6 & 32 &  0.246 &   13.3 & 0.7 -- 1.0 $\arcsec$  seeing \\
2014 May 15 & Polarization & $K1$ & 59.6 & 16  &  0.306 &   11.4 & 0.7 -- 1.0 $\arcsec$ seeing  

\enddata
\end{deluxetable*}

The Gemini Planet Imager (GPI) was designed to overcome the contrast problem inherent in the detection of circumstellar material within $\sim 1.5 \arcsec$ from their host stars.  GPI combines an advanced adaptive optics (AO) system, an apodized coronagraph, and an IR integral field spectrograph with both spectral and polarimetric modes \citep{2014PNAS..11112661M, 2014SPIE.9147E..1KL,2015ApJ...799..182P}. 

PDS 66 (MP Muscae) is one of the closest T Tauri stars. It was identified as part of the Pico Dos Dias Survey \citep{1992AJ....103..549G}. 
\citet{2002AJ....124.1670M} 
classified PDS 66 as a member of the Lower Centaurus Crux (LCC) subgroup with a mean age of $17 \pm 1$ Myrs. \citet{2002AJ....124.1670M} list PDS 66 as a K1 spectral type star with a kinematic parallax distance of $86^{\,+8}_{\,-7}$ pc and age estimates ranging from 7 -- 17 Myrs. 
PDS 66 was the \textit{only} cTTS found in their sample of over 100 pre-main-sequence stars.  
\citet{2008hsf2.book..757T} first suggested that PDS 66 is more likely a member of the $\epsilon$ Cha Association. \citet{2013MNRAS.435.1325M} reinvestigated the membership of PDS 66 and found that the proper motion is more consistent with $\epsilon$ Cha (age: 5 -- 7 Myrs, kinematic distance: $101 \pm 5$ pc). 
The membership of PDS 66 remains somewhat uncertain between LCC and $\epsilon$ Cha. Given that the disk properties of PDS 66 appear to be inconsistent with the LCC, and the younger age of $\epsilon$ Cha is below the typical disk dissipation timescale of 10 Myr \citep{2001ApJ...553L.153H}, in this paper, we adopt the age and distance appropriate for eps Cha.

The PDS 66 disk is in an interesting evolutionary stage. The spectral energy distribution (SED) lacks signs of large-scale evolution or an inner clearing. Based on the 1.2 mm continuum flux \citet{2005AJ....129.1049C} estimated a total dust mass of $5.0 \times 10^{-5} M_{\odot}$. \citet{2005A&A...431..165S} modeled the PDS 66 mid-infrared spectra and SED 
and infer an inner disk radius consistent with the dust sublimation radius of 0.1 AU. However, \citet{2009ApJ...697.1305C} compare the PDS 66 SED to the median Taurus SED and find a flux decrement between 4-20 microns, indicating a partial clearing of material in the disk. 
CO measurements by \citet{2010ApJ...723L.248K} show a molecular gas disk extending out to 120 AU with a lower limit for the gas mass of $9.0 \times 10^{-6} \, M_{\odot}$.  
Although uncertain, this suggests a lower gas-to-dust ratio limit of $\ge 0.2$. 
Even though the accretion rate inferred for PDS 66 is small for a cTTS \citep[estimates range from $5 \times 10^{-9}$ to  $1.3 \times 10^{-10} \, M_{\odot}/yr$][]{2007ApJ...663..383P,2013ApJ...767..112I}, the implied accretion timescale is short, $< \, 10^{5}$ yrs. (based on the disk mass inferred from CO).


The PDS 66 circumstellar disk was first resolved in \textit{HST}/NICMOS imaging by \citet{2009ApJ...697.1305C}. They detected a disk with an outer radius of 170 AU, and an inclination of $32 \degr \pm 5 \degr$. 
The authors also provide evidence for grain growth through an analysis of the spectral energy distribution.
Likewise, \citet{2008ApJ...683..479B} obtained Spitzer spectroscopy ($8 - 13 \, \mu m$) and found that the dust grain properties are well fit by a model consisting of amorphous olivine and  pyroxene with average particle sizes of a few microns. 
\citet{2014AJ....148...59S} obtained deep \textit{HST}/STIS coronagraphy showing consistent geometry, albeit with detection of faint halo extending out to beyond 520 AU.

PDS 66 was observed during the commissioning of GPI to test instrument performance on a typical bright, nearby disk. Our GPI observations are described in Section 2. The morphology of the PDS 66 disk as seen in polarized light with GPI is discussed in Section 3. We place limits on our sensitivity to planetary companions in Section 4. Section 5 discusses the results.

\section{Observations}

\begin{figure*}[t!]
\begin{center}
\includegraphics[scale=0.45]{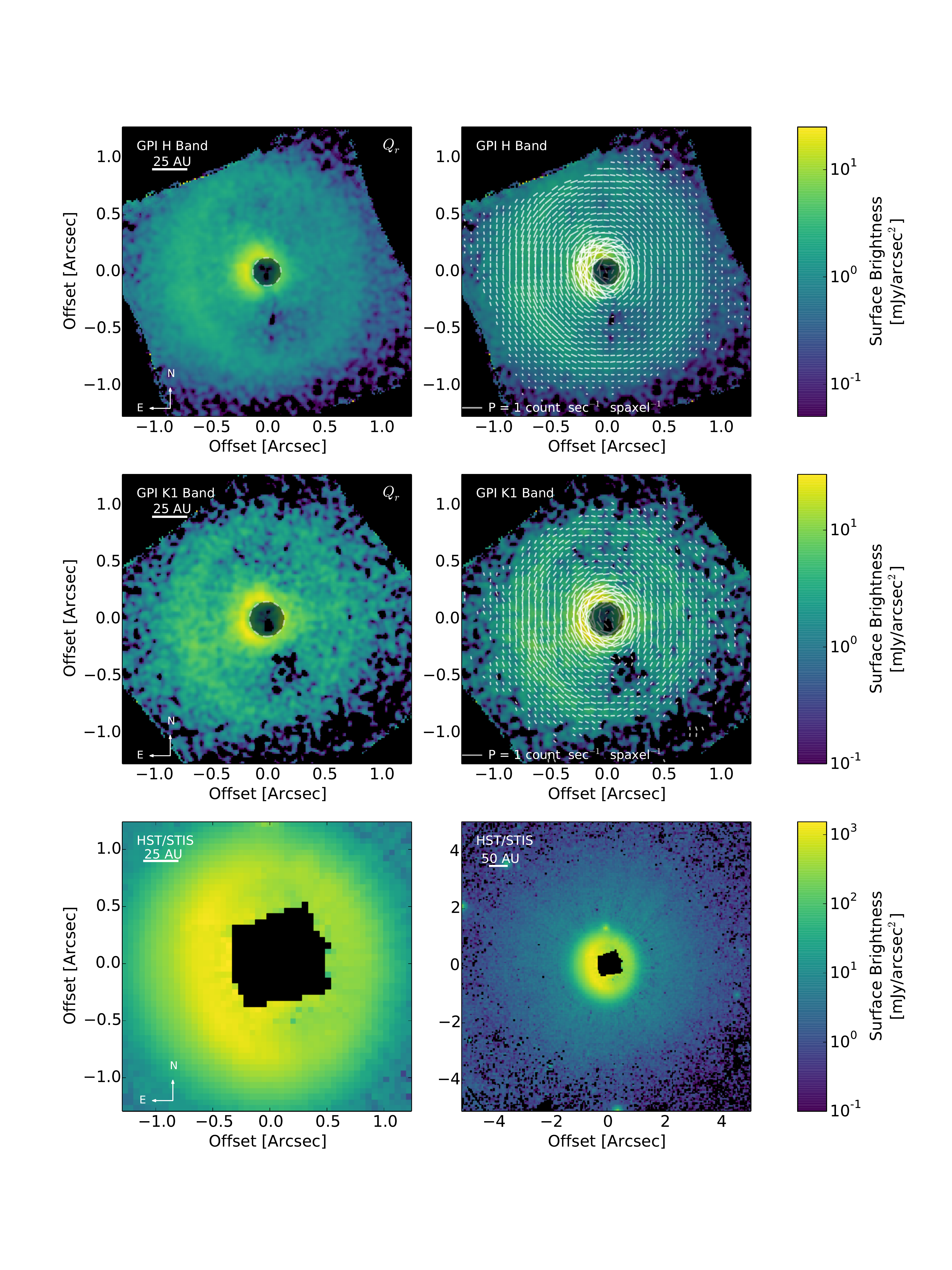}
\caption{Polarimetry data for PDS 66 in \textit{H} band (top) and \textit{K1} band (middle), and white light optical STIS data at two spatial scales for comparison (bottom; data from Schneider et al. 2014). Polarized intensity is shown on the left for the GPI data, while the right panels show the same polarized intensity over-plotted with polarization vectors. The vector orientation gives the position angle for the polarized electric field. Grey inner regions represent the coronagraphic spot size.   \label{fig1}}
\end{center}
\end{figure*}

Coronagraphic imaging polarimetry and spectroscopy of PDS 66 were obtained in 2014 May (Table 1). The GPI Integral Field Spectograph (IFS) has a plate scale of $0.014$ arcseconds/pixel, a FOV of 2.8 X 2.8 arcsec$^{2}$, and an angular resolution of $\sim 0.05''$ in H band \citep{2014PNAS..11112661M, 2014SPIE.9147E..1KL}.
Data were reduced using the GPI Data Reduction Pipeline; see \citet{2014SPIE.9147E..3JP} and references therein. Data were obtained in high wind conditions with the AO system operating at 500Hz. 
For the spectral mode data, the raw frames were dark subtracted, corrected for bad pixels, destriped to correct for variations across read-out channels, and Fourier filtered to remove microphonics noise. A wavelength calibration using arc lamp data taken before the sequence is used to convert the raw images to 3D spectral datacubes \citep{2014SPIE.9147E..7HW}. 
The location of the star behind the coronagraphic mask was measured from the satellite spots \citep{2006ApJ...647..620S, 2014SPIE.9147E..55W} and the data were corrected for spatial distortion. 

In polarimetry mode, frames are taken in sets of four different half-wave plate rotations and combined to form Stokes cubes with slices \textit{I, Q, U,} and \textit{V}. Data are dark subtracted, destriped, 
and a thermal sky background is subtracted (in \textit{K1} band only). The individual frames are converted into two orthogonal polarization states using a spot location calibration file that has been corrected for elevation-induced flexure. 
Each cube is divided by a low pass filtered flat field to correct for low frequency variations \citep{2015arXiv150804787M}. The mean stellar polarization and instrumental polarization are subtracted and the polarization pairs are cleaned via a double difference algorithm \citep{2015ApJ...799..182P}. The satellite spots are again used to determine the location of the occulted star and to calibrate the flux of the disk using a conversion factor of $1$ ADU coadd$^{-1}$ s$^{-1} = 7.4 \pm 2.6$ mJy arcsec$^{-2}$ in \textit{H} band and $31 \pm 10$ mJy arcsec$^{-2}$  in \textit{K1} band \citep{2015ApJ...815L..14H}. 


Figure 1 shows the \textit{H} and \textit{K1} band polarimetry for PDS 66 with the Stokes vectors giving the orientation. 
Here the Stokes parameters have been transformed to radial Stokes parameters \citep{2006A&A...452..657S}. The $+Q_{r}$ image contains the polarization oriented in the tangential  direction in the disk, $-Q_{r}$ contains the radial polarization and $U_{r}$ contains the polarization oriented $\pm 45 \degr$ from $Q_{r}$. For an optically thin disk, the $U_{r}$ image should contain no polarized flux from the disk and can be treated as a noise map. 
For an optically thick disk like PDS 66, multiple scattering events can result in non-negligible brightness, at a few \% of the $Q_{r}$ signal for low-inclination disks \citep{2015A&A...582L...7C}. Given this small amplitude, we adopt
the $U_{r}$ channel as a measure of our errors, recognizing that the contribution of both noise and potential signal renders it a conservative estimate.

The spectral mode data were PSF-subtracted using the pyKLIP software \citep{pyKLIP}\footnote{https://bitbucket.org/pyKLIP/pyklip}. pyKLIP combines both Spectral Differential Imaging (SDI: for spectral mode data) and Angular Differential Imaging (ADI) using the Karhunen-Loeve Image Projection (KLIP) algorithm \citep{2012ApJ...755L..28S}. 
Due to the face-on nature of the disk, recovery of the total intensity is difficult via ADI. 
We leave forward modeling of the disk's total intensity surface brightness, and
calculation of the polarization fraction, to future work.

\begin{figure*}
\begin{center}
\includegraphics[scale=0.4]{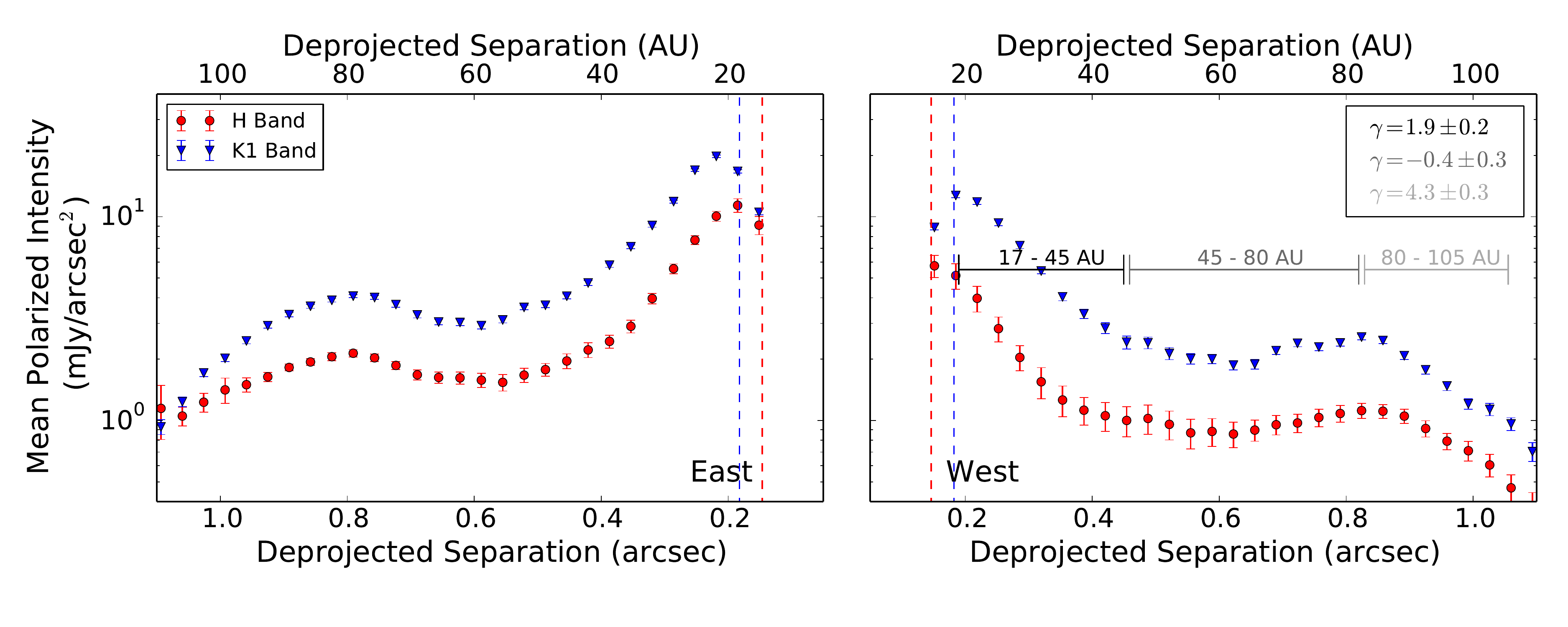}
\caption{Radial brightness profile of the tangential polarized intensity for \textit{H} and \textit{K1} bands for the East (left) and West (right) sides of the disk. Vertical dashed lines indicate the outer edge of the coronagraphic spot in \textit{H} and \textit{K1} bands. Fits to the power law slopes ($\gamma$) are given in the legend (E/W slopes agree). Error bars are drawn from the $U_{r}$ error maps. The profile shows the bright inner ring of material and a peak at $\sim 0.8\arcsec$ (80 AU) corresponding to the outer ring. }
\end{center}
\end{figure*}

\section{Disk Morphology}

The GPI data reveal a bright disk interior to a more diffuse disk extending to an outer ring, and an azimuthal asymmetry indicative of interesting structure close in to the central star (Figure 1).
 We also show the STIS data provided by \citet{2014AJ....148...59S} to illustrate the fainter outer halo outside the field of view of GPI. 
 The inner disk likely extends from the sublimation radius to the change in the power law slope at 45 AU.
 The region between the inner disk and outer ring (45 -- 80 AU) is not entirely cleared, as evidenced by the azimuthal orientation of the polarization vectors.
 

We fit an ellipse to the brightness contours in the outer disk ring using the constrained, linear, least squares method described in \citet{1998proc}. We find a position angle for the disk major axis of $10\degr \pm 3\degr$ E of N, an axial ratio of $0.86 \pm 0.02$, and a disk inclination of  $31\degr \pm 2 \degr$ from a face on viewing geometry. These values agree well with the STIS results \citep[minor:major axial ratio $0.889 \pm 0.026$, inclination $27.3 \pm 3.3$ degrees:][]{2014AJ....148...59S}. 
We measure no stellocentric offset to within 30mas, consistent within errors with the offset in the STIS observations of $33 \pm10$ mas \citep{2014AJ....148...59S}. Low SNR in the satellite spots of these observations limits our knowledge of the obscured star's location to within $\sim 2$ pixels.

We deproject the disk and calculate a radial brightness profile (Figure 2) separately for the East and West sides of the disk. 
Note that the peak in surface brightness is slightly offset from the edge of the coronagraphic mask. This is likely due to a lower throughput from an instrumental effect rather than a decrease in the surface density of the disk \citep[See also][]{2015ApJ...815L..26R}. 
The East side of the disk is brighter in both total intensity (STIS/NICMOS) and polarized intensity (GPI). Since we expect the dust particles in the disk to be predominately forward scattering, we conclude that the East side is the nearer side. 
We fit power laws $\propto r^{- \gamma}$ to the surface brightness profile in the inner disk, the central region, and the outer ring (see Figure 2). The power law slope in the inner disk is consistent with an optically thick, gas-rich disk. For the outer component (80 -- 105 AU), the GPI power law fit agrees well with the STIS and NICMOS result \citep{2009ApJ...697.1305C}.

After correcting for extinction ($A_{V} = 0.7 \pm 0.2$ mag; \citet{2009ApJ...697.1305C}) and stellar color (assuming a K1 spectral type with intrinsic H-K = 0.14), the azimuthally-averaged apparent color of the disk is \textit{H-K} $= 0.45 \pm 0.17$ in polarized intensity, implying that the dust in the disk is $\sim$ 50\% more effective at reflecting \textit{K1} band light than \textit{H} band light. 
In \textit{H} band, the East side of the disk is 2.1 times brighter than the West side, while the East side is only 1.6 times brighter than the West in  \textit{K1} band. The E/W flux ratio is much lower than seen in total intensity in the visible \citep{2014AJ....148...59S}, which suggests either more isotropic scattering and/or a high polarization fraction on the (fainter) W side.

A region within the south side of the disk appears depleted in polarized intensity in both \textit{H} and \textit{K1} bands. Figure 3 shows the azimuthal brightness variations for two disk annuli (35 -- 50 AU and 70 -- 90 AU) computed from the mean and standard deviation in $12 \degr$ wedges. In the 35 to 50 AU region, there is a $\sim$ 35 \% decrease in the surface brightness from PA $160\degr - 220\degr$ (measured E from N). \citet{2014AJ....148...59S} also saw brightness asymmetries of $\sim$ 30\% between two epochs of data spaced three months apart. Though at a different parallactic angle, the drop in brightness subtends approximately the same angular fraction of the disk.

\begin{figure*}
\begin{center}
\includegraphics[scale=0.4]{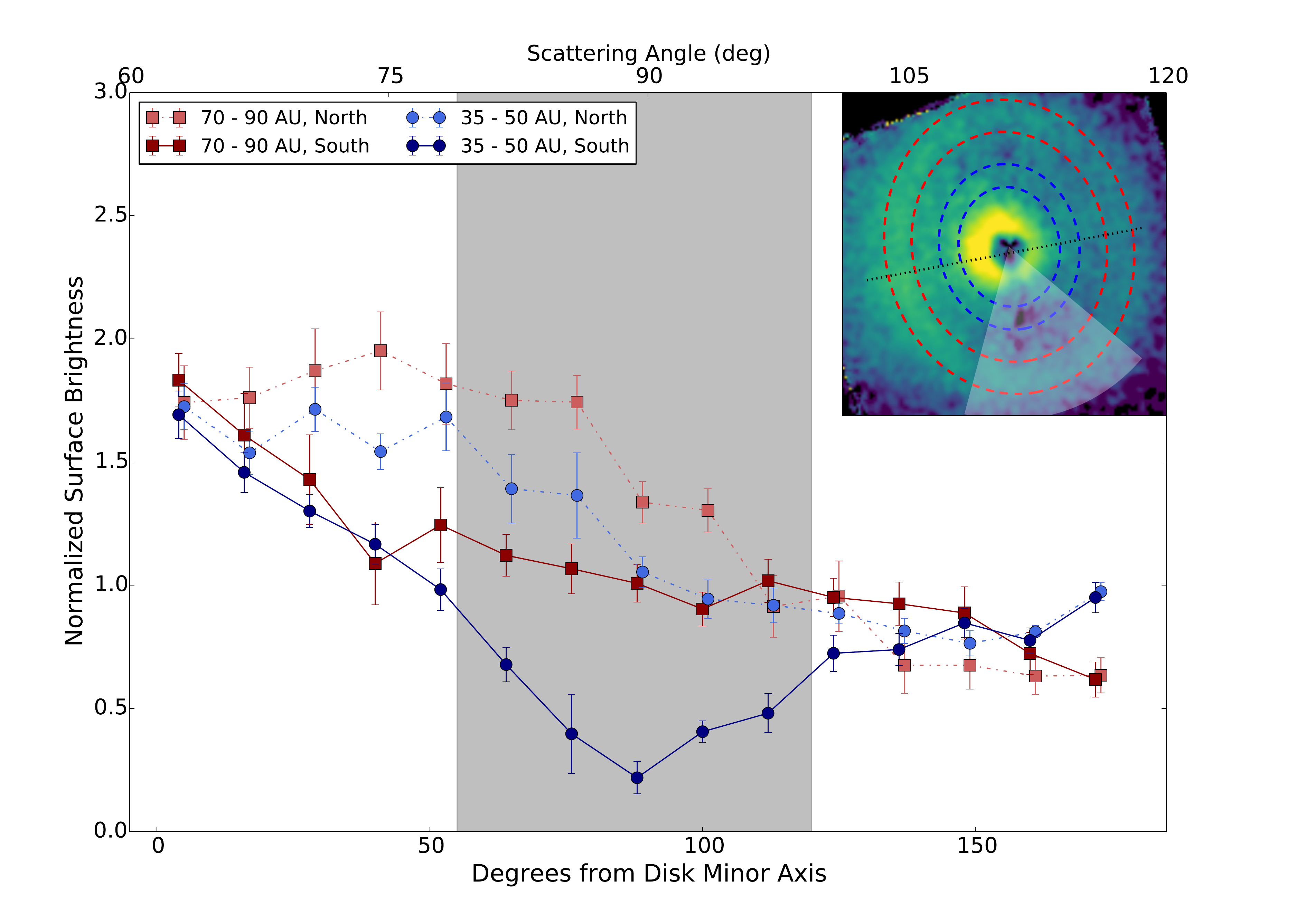}
\caption{ The azimuthal variation of the median polarized intensity as measured for the \textit{H} band GPI data in an annulus in the gap from 35 to 50 AU and for the bright ring from 70 to 90 AU. The North and South polarized intensities are plotted separately to emphasize the drop in flux seen on the South side of the disk. Polarized intensity values have been normalized to the mean separately for each ring. The inset shows the \textit{H} band image (North up). The dashed blue/red lines represent the annulus used to measure the azimuthal variation. The black dotted line gives the location of the disk minor axis. The gray shaded region corresponds to the shaded inset wedge. }
\end{center}
\end{figure*}

\section{Planetary Companion Limits} \label{companions}

Our spectral data constrain planetary companions of a given mass and age. 
We compute a $5\sigma$ contrast curve assuming a methane dominated planetary spectrum. We achieve a contrast of $\sim 10^{-5}$ outside of 0.3$\arcsec$ and $\sim 2 \times 10^{-6}$ outside of 0.4$\arcsec$. We detect no planetary candidates, but we recover a bright source in the north at 50$\sigma$, which was previously confirmed as a background source \citep{2014AJ....148...59S, 2009ApJ...697.1305C}. 

Planet sensitivities are calculated following \citet{2010ApJ...717..878N} and \citet{2008ApJ...674..466N} (Figure 4). The contrast curve is used to set companion brightness limits with radius. The brightness of a planet with a given mass and age are set by the hot start evolutionary tracks of \citet{2003A&A...402..701B}. 
For an age of 7 Myr and a distance of 100 pc ($\epsilon$ Cha membership) there is a 90 \% confidence that we would have detected a 8 $M_{Jup}$ planet at $\sim$ 20 AU or a 3 $M_\mathrm{Jup}$ planet outside of 40 AU. At 17 Myrs and 86 pc (LCC membership), the 90 \% confidence limits increase to a 10 $M_\mathrm{Jup}$ planet at $\sim$ 20 AU. 
Planetary companions may exist, but lack a methane absorption feature, or could be low-mass enough to remain hidden below the opaque disk surface. 

\section{Discussion}





PDS 66 joins the class of pre-transitional disks \citep{2010ApJ...717..441E} with an optically thick inner disk separated from an outer disk by a dip in surface brightness around $0.5 \arcsec$ that could indicate a partial clearing of the disk. The gap/ring structure observed in our GPI data, combined with the detection of orbiting CO \citep{2010ApJ...723L.248K} confirm that PDS 66 closely resembles the V4046 Sgr and TW Hya systems. All are nearby cTTS that have retained their molecular gas to late ages and show multi-ringed structures. 
GPI polarimetry was used to confirm the presence of scattering dust in the gaps of the V4046 Sgr multi-ringed structure \citep{2015ApJ...803L..10R}. 
TW Hya is multi-ringed with partially filled gaps as well \citep{2013ApJ...771...45D,2015arXiv151201865R}.  

 If the disk is optically thick, the ring/gap structure is a result of a variation in the disk surface that could be caused by a change in the surface density, the local scale height, or the dust properties of the sub-micron sized grains in the disk. Here we discuss possible sources for a change in the disk surface properties: 

\begin{enumerate}[{(a)}]

\item \textit{Gap Opening Planets:} A planet/(s) in the low surface brightness region could induce a gap in the dust disk and deplete the gas \citep{2015ApJ...809...93D}. Dust filtration is efficient at piling up larger dust particles (mm-sized) into a ring at the pressure bump outside of a gas gap \citep{2015ApJ...806L...7Z}, while smaller grains (responsible for scattered light) could still populate the gap. Given the observed width ($\sim 35$ AU) and the shallow depth (ring:gap = 1.4), this is most likely a planetary system with several sub-jupiter mass planets.Ê

\item \textit{Disk Shadowing:} A scale height enhancement in the inner part of the disk shadows the outer disk, until the flaring of the disk eventually brings the disk surface above the penumbra. \citet{2015ApJ...810....6D} find that a puffed up inner wall can create a three part broken power law in the radial brightness profile, as seen in the GPI data. A shadow cast out to 80 AU would require a flat disk and/or a low flaring exponent.

\item \textit{Dust Particle Properties:} A localized change in the dust properties would change the opacity of the disk. Dust settling due to grain growth could induce a change in the scale height, which would change the height of the scattering surface relative to the disk midplane, producing the bright ring. Gaps in the HL Tau disk have been ascribed to the effects of snowlines \citep{2015ApJ...806L...7Z}. However, given the large radius of the observed ring (80 AU), this seems unlikely. 

\end{enumerate}

\begin{figure*}[t!]
\begin{center}
\includegraphics[scale=0.55]{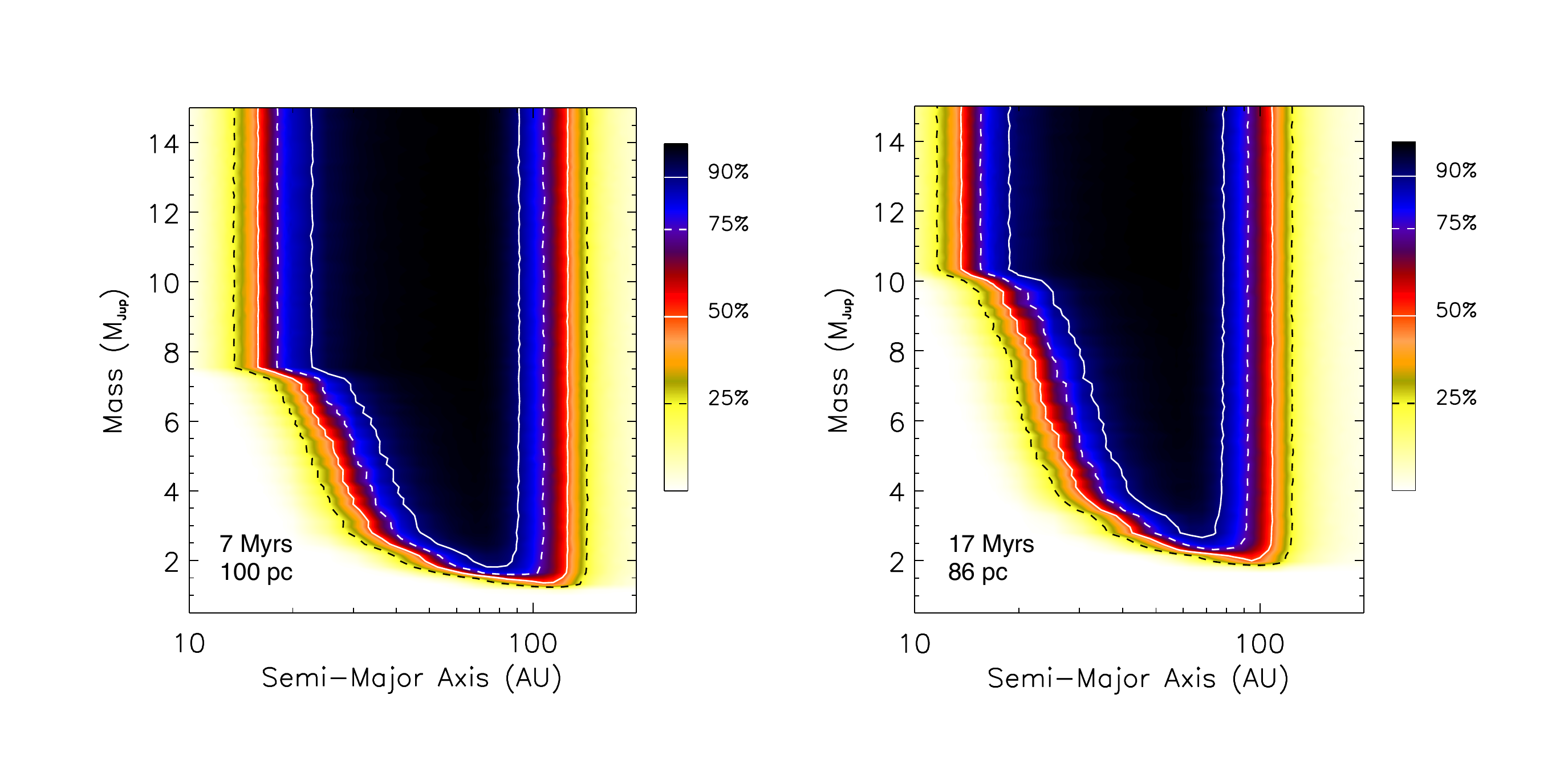}
\caption{Companion sensitivity as a function of separation and mass for membership in $\epsilon$ Cha (Left) and the LCC (Right) with ages and distances as shown.}
\end{center}
\end{figure*}

For an inclined disk that is optically thick vertically and axisymmetric, a ring with a higher surface height would appear as an offset structure relative to the central star \citep{2006Sci...314..621L}. Combining the offset measured in the STIS image with the ring radius, we infer that the scattering in the ring occurs 4 AU above the disk midplane. The expected scale height for gas in vertical hydrostatic equilibrium at the location of the ring is about 4--5\,AU (assuming $T_{eff}=5000$\,K, $L=1.1 L_\star$ and $T_{80 {\rm AU}}=10 - 15$\,K), i.e. similar to the height where scattering occurs. In optically thick disks, the disk surface is typically located 2--4 times higher than the gas scale height \citep[e.g.,][]{1999ApJ...527..893D}. This suggests that the PDS 66 disk is flatter and/or less flared than primordial disks, i.e., possibly significantly settled as was originally suggested by \citet{2009ApJ...697.1305C}. A flattened disk could favor the "shadowing" scenario above, but only a more complete SED+image modeling effort can confirm this.

From this dataset, no clear conclusions can be drawn on the origin of the gap + ring structure. ALMA dust continuum observations would help distringuish between the scenarios above. For scenario (a), we would expect to see a significant pile up of mm-sized grains right outside the NIR ring, due to the dust filtration effect, which would generate at least a factor of $\sim 10$ or higher in continuum flux. In scenario (b), the shadowed region would have a slightly lower temperature, which would result in less flux in the optically-thin mm continuum as well, though only on the order of $\lesssim 50$\%. ALMA gas observations may be able to detect gas depletion in the scenario (a), however given the shallowness of the gap this may not provide sufficient contrast between shadowed and unshadowed regions.

We detect an azimuthal departure from axisymmetry, seen as a dimmer region in the disk's southern side at around a 40 AU radius. 
\citet{2014AJ....148...59S} observed that the east/west asymmetry of the disk is variable on timescales as short as three months \citep{2014AJ....148...59S}. Since this is much shorter than the dynamical timescales at the relevant orbital separations, \citet{2014AJ....148...59S} hypothesized that the changes could be due to either time-variable shadowing from material in the inner disk hidden behind the coronagraphic mask, or localized accretion hot spots on the stellar photosphere. It is possible that the azimuthal asymmetry seen in the GPI data (at $\sim$ 40 AU) is due to such an effect, rotated around to affect the illumination over a different range of position angles. 
Density enhancements in the disk caused by accreting protoplanets might cast shadows on the outer regions of the disk, though the shadowed areas predicted by simulations for planets as massive as 50 $M_{Earth}$ are only $\sim 7 \, AU^{2}$ \citep{2009ApJ...700..820J}.  
Alternatively, cold spots on the stellar surface which are darker due to magnetic suppression of convection typically cover 5 - 30\% of the stellar surface and could cause an azimuthal modulation of the stellar illumination incident on the outer disk on stellar rotation timescales \citep{2015A&A...581A..66V}.

If the azimuthally variable disk surface brightness distribution is due to nonuniform brightness on the stellar surface, it will change on timescales of the rotation period (5~days). If instead it is due to material orbiting at the estimated inner radius (10.5~days at 0.1~AU) or embedded in the bright inner ring the shadowing will vary over a longer period.
More data is needed to elucidate the timescales of the azimuthally variable disk surface brightness.


\textbf{Acknowledgements:} We thank the referee, Joel Kastner, for his advice that helped strengthen this paper. We acknowledge financial support of Gemini Observatory, the NSF Center for Adaptive Optics at UC Santa Cruz, the NSF (AST-0909188; AST-1211562; AST-1413718), NASA Origins (NNX11AD21G; NNX10AH31G), the University of California Office of the President (LFRP-118057), and the Dunlap Institute, University of Toronto. This work is supported by the National Science Foundation Graduate Research Fellowship under Grant No. DGE-1232825. Portions of this work were performed under the auspices of the U.S. Department of Energy by Lawrence Livermore National Laboratory under Contract DEAC52-07NA27344 and under contract with the California Institute of Technology/Jet Propulsion Laboratory funded by NASA through the Sagan Fellowship Program executed by the NASA Exoplanet Science Institute. This study is based in part on observations made with the NASA/ESA Hubble Space Telescope (program GO 12228), obtained at the Space Telescope Science Institute, which is operated by the Association of Universities for Research in Astronomy, Inc., under NASA contract NAS 5Ð26555. 
Based on observations obtained at the Gemini Observatory.

\clearpage

\end{document}